\documentclass{article}
\usepackage{spconf,amsmath,graphicx}

\usepackage{graphicx}
\usepackage{caption}
\usepackage{subcaption}
\usepackage{multirow}
\usepackage{color}
\usepackage{xcolor}
\graphicspath{{figures/}{pictures/}{images/}{./}} 
\DeclareMathOperator{\boldx}{\mathbf{x}}

\usepackage{soul}

\title{Synthetic Wave-Geometric Impulse Responses for Improved Speech Dereverberation}
\name{Rohith Aralikatti, Zhenyu Tang, Dinesh Manocha}
\address{
  University of Maryland, College Park}

\begin{document}

\maketitle

\begin{abstract}
We present a novel approach to improve the performance of learning-based speech dereverberation using accurate synthetic datasets. Our approach is designed to recover the reverb-free signal from a reverberant speech signal. 
We show that accurately simulating the low-frequency components of Room Impulse Responses (RIRs) is important to achieving good dereverberation. We use the GWA dataset \cite{gwa} that consists of synthetic RIRs generated in a hybrid fashion: an accurate wave-based solver is used to simulate the lower frequencies and geometric ray tracing methods simulate the higher frequencies.
We demonstrate that speech dereverberation models trained on hybrid synthetic RIRs outperform models trained on RIRs generated by prior geometric ray tracing methods on four real-world RIR datasets. 
\end{abstract}

\section{Introduction}

Speech processing applications tend to exhibit lower performance in the presence of reverberation. Several research works have documented this phenomenon for standard tasks such as automatic speech recognition (ASR)~\cite{ko2017study} and speech separation~\cite{aralikatti2021improving}. The degradation in model performance can be attributed to a few factors. These include reverberant distortions that cause a reduction in signal intelligibility and the distributional mismatch between reverb-free training data and noisy reverberant test data.

A straightforward way to reduce such mismatch is to collect more training data similar to the test data in specific environments. 
However, speech data that includes diverse environmental effects can be difficult to collect in the real world. 
One common approach to improving ASR model robustness under real-world reverberation is by artificially converting annotated clean speech to reverberant speech data and re-training the ASR model. 
This is typically performed by convolving clean speech data with room impulse responses (RIRs) corresponding to different environments. 
An RIR encodes how an environment modifies sounds for a specified source and receiver location pair, based on propagation of sound in the environment. 
Alternatively, without re-training the ASR model, one can utilize a speech dereverberation model as a pre-processing layer to remove environmental reverberation for a generic ASR model. 
Speech dereverberation has been shown to improve the performance of various tasks such as ASR and speaker verification~\cite{kothapally2020skipconvnet}. 

Several recent works have proposed learning-based solutions for speech dereverberation ~\cite{kothapally2020skipconvnet,ernst2018speech,fu2019metricgan}. 
Training robust speech dereverberation models requires the existence of large high-quality RIR datasets to capture all possible variations present in real environments. However, capturing real-world RIRs is a time-consuming process that requires professional or expensive hardware~\cite{farina2000simultaneous}. There is a large body of work on generating synthetic RIRs using sound simulation algorithms. These sound propagation methods~\cite{liu2020sound} can be broadly categorized into two main categories: geometric or ray tracing-based methods \cite{allen1979image,taylor2012guided} and wave-based or numeric methods \cite{botteldooren1995finite,thompson2006review}. Geometric methods are based on a fundamental assumption that sound waves travel as rays and undergo geometric reflections  from object surfaces. However, this assumption is not valid for low-frequency sounds (e.g., $<1000Hz$). In these cases, the wavelength of the sound is comparable to the size of common objects present in daily life and significant non-linear wave effects can occur. We need to use accurate numeric solvers of wave-equations to simulate these sound effects.

\noindent \textbf{Main Contributions:} 
Some novel components of our work include: 
\begin{itemize}
 \item We show that accurately modeling the low-frequency components of RIRs is important to obtain better dereverberation.
     \item We show that the hybrid approach that combines wave and geometric methods for simulating RIRs improves Speech-to-Reverberation Modulation energy Ratio (SRMR) by $8.3\%$ (average) relative to the synthetic RIRs generated using geometric methods. 
\end{itemize}


\section{Related Work}

\subsection{Speech Dereverberation}
\vspace{-2mm}
Several approaches have been proposed to solve speech dereverberation.
Weighted Prediction Error (WPE) \cite{nakatani2010} is one such method that uses variance delayed linear prediction to estimate the late reverberation present in the signal. 
Ernst et al.~\cite{ernst2018speech} proposed a fully convolutional network (FCN) for speech dereverberation. Koothapally et al. proposed SkipConvNet~\cite{kothapally2020skipconvnet}, an improved FCN with skip connections between the encoder and decoder layers. MetricGAN~\cite{fu2019metricgan} utilized a GAN-based approach where the generator attempted to maximize speech metrics such as Perceptual Evaluation of Speech Quality (PESQ) and Speech-to-Reverberation Modulation energy Ratio (SRMR). 
In the works above, the dereverberation models have been trained on reverberant data generated from a limited quantity of recorded RIRs. There has not been much work done on utilizing different synthetic RIR generation methods to train dereverberation models that can generalize well to a large variety of acoustic environments at test time.
\vspace{-2mm}
\subsection{Synthetic Data for Speech Processing}
\vspace{-2mm}
There is considerable work on generating synthetic datasets and RIRs to improve the accuracy of learning-based speech processing applications.
Many speech processing works have used geometric acoustic simulation methods to generate synthetic training data.
One of the most widely used methods is the \emph{image-source method (ISM)}~\cite{allen1979image}, which assumes pure specular reflections for sound rays and builds ``image sources'' by mirroring the source according to known planar surfaces in the scene. 
The use of ISM simulation has resulted in significant improvement in speech recognition performance for some applications~\cite{ko2017study}. Other methods improve the accuracy of RIRs
by also modeling diffuse sound reflections on rough surfaces. 
These include path tracing methods~\cite{taylor2012guided} based on efficient Monte Carlo path tracing~\cite{kajiya1986rendering} and beam or frustum tracing methods~\cite{funkhouser1998beam}.
These \emph{geometric acoustic (GA)} methods can generate more accurate RIRs than ISM simulations and have been beneficial for speech processing benchmarks~\cite{tang2020improving}. Wave-based simulation methods result in the most realistic RIRs. They are described in more detail in Section. \ref{ssec:wave}. 

\

\begin{figure}[tb]
\centering
\begin{subfigure}[b]{0.49\linewidth}
  \centering
  \includegraphics[width=\textwidth]{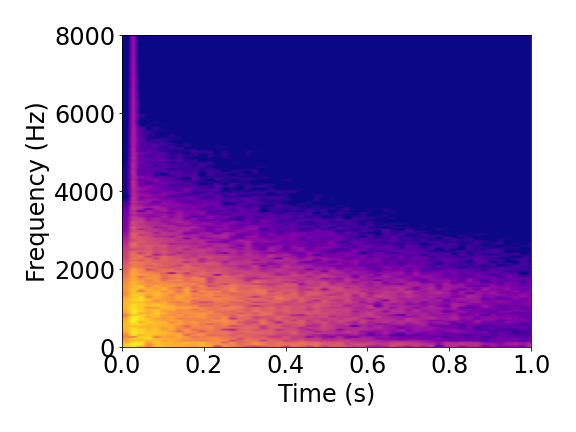}
  \caption{FDTD RIR}
  \label{fig:FDTD} 
\end{subfigure}
\begin{subfigure}[b]{0.49\linewidth}
  \centering
  \includegraphics[width=\textwidth]{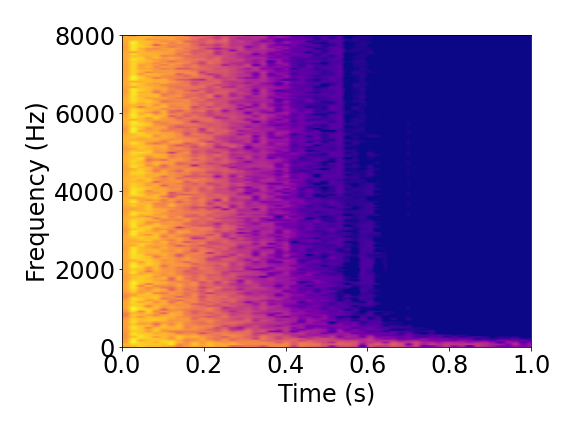}
  \caption{Geometric RIR}
  \label{fig:Geo}
\end{subfigure}
\begin{subfigure}[b]{0.49\linewidth}
  \centering
  \includegraphics[width=\textwidth]{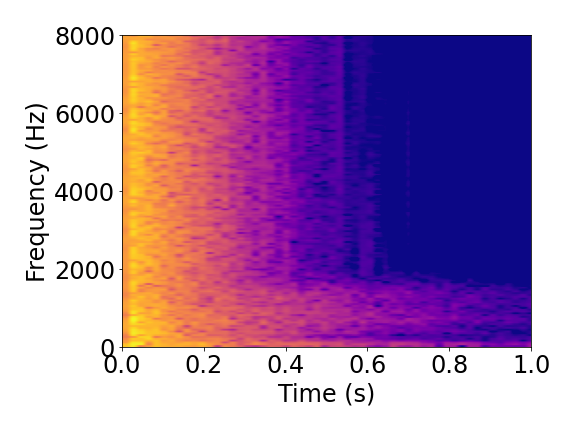}
  \caption{Hybrid RIR}
  \label{fig:hybrid}
\end{subfigure}
\begin{subfigure}[b]{0.49\linewidth}
  \centering
  \includegraphics[width=\textwidth]{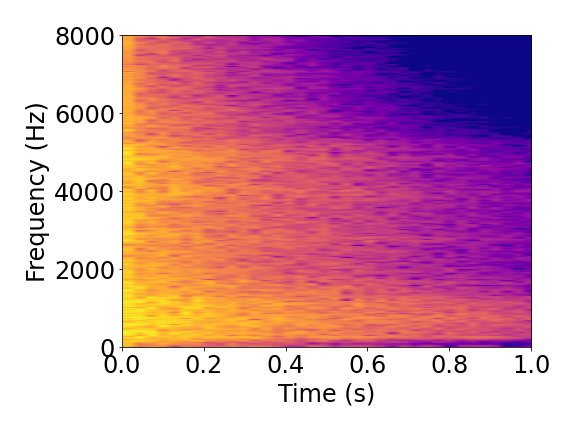}
  \caption{Real RIR}
  \label{fig:real}
\end{subfigure}

\caption{Comparison of spectrograms for real and synthetic RIRs (darker area has lower energy). The FDTD simulation is used to compute the lower frequency components in the hybrid spectrogram, which matches the FDTD spectrogram. The synthetic RIRs in (a)(b)(c) are all generated in the same conditions and environment. A real RIR (d) from the MIT IR dataset is chosen for qualitative comparison since the synthetic RIRs do not aim to match this particular scene. We observe that the low frequency components in the real RIR have stronger energy while the geometric RIR has a relatively flat distribution, and the hybrid RIR has a similar energy distribution to that of the real RIR. } 
\end{figure}
\vspace{-0mm}



\vspace{-5mm}

\section{Training with Low-Frequency Speech}
\vspace{-2mm}
In speech processing, the fundamental frequency of human speech is generally below 250Hz~\cite{traunmuller1995frequency}, although the usual sampling rate is above 8,000Hz. It is also evident that low-frequency components in speech play an important role for speech intelligibility~\cite{brown2009low} and contribute strongly to tone recognition in various languages~\cite{luo2006contribution}. 
In addition, some speech processing models show that the convolution layers learn directly from speech data to emphasize low-frequency speech regions (below 1000Hz)~\cite{muckenhirn2018towards}. Tang et al. \cite{tang2020low} show that compensating the low-frequency simulation components by matching the equalization distributions of synthetic RIRs with those of real RIRs reduces the Word Error Rate (WER) of ASR systems in certain conditions.  

These findings show that low-frequency speech signals should be modeled accurately and that they could have significant impacts on the models trained using such accurate data. 
Most prior methods for generating synthetic RIRs and datasets (see Section 2.3) only use geometric acoustic simulation methods.
some important low-frequency room acoustic effects or components are often missing in GA simulations. 
One obvious caveat of geometric methods is sound diffraction, in which a geometric sound ray can easily be blocked  by a small obstacle between the source and the listener. 
However, such results clearly deviate from real-world physics, as the sound wave tends to bend over small obstacles. Our goal is to model these effects accurately.


\subsection{Wave-Based Acoustic Simulation}
\label{ssec:wave}
We use wave-based methods to generate accurate RIRs.
A scalar acoustic pressure field, $P(\boldx, t)$, satisfies the  wave equation
\begin{equation}
\label{eq:waveequation}
\frac{ \partial^2 P(\boldx, t)}{\partial t ^2 } - c^2\nabla ^2 P(\boldx, t) = f(\boldx, t),
\end{equation}
where $c$ is the speed of sound, $\boldx$ is the 3D coordinate, and $f(\boldx, t)$ is the forcing term, usually representing some driving source signal. 
An RIR can be obtained by setting $f(\boldx, t)$ to an impulse signal at a source location $\boldx_s$, fixing $P(\boldx, t)$ at the receiver location $\boldx_r$ and extracting its time-varying component. 
The wave equation can be solved numerically using different solvers: finite-difference time domain (FDTD)~\cite{botteldooren1995finite}, finite-element (FEM) method~\cite{thompson2006review}, the adaptive rectangular decomposition (ARD) method~\cite{raghuvanshi2009efficient}, etc. 
Wave-based methods can correctly model sound diffraction for non-line-of-sight (NLOS) source and listener pairs, which are common in real world scenarios. 
Furthermore, room shape can add extra resonant frequencies that modify the RIR at some locations due to standing waves. This phenomenon is more dominant at low frequencies and is called \emph{room modes}, which is also modeled by wave-based methods but not geometric methods. 
\vspace{-3mm}
\subsection{Datasets}
\vspace{-2mm}
Despite its high accuracy in the low-frequency range, wave acoustic simulation can be time consuming..
We use pre-computed RIRs that combine the low-frequency components from wave-based simulations and high-frequency components from geometric simulations. 
Specifically, we utilize RIRs from the GWA dataset\footnote{https://gamma.umd.edu/pro/sound/gwa}~\cite{gwa}. 
GWA contains 2 million realistic RIRs simulated in more than 5,600 rooms with diverse real-world material configurations. 
In order to generate synthetic RIRs, source and listener locations are uniformly sampled in each 3D room model with furniture and pre-assigned acoustic materials. 
Wave acoustic results are obtained with an FDTD solver up to 1,400Hz, as that is the chosen cutoff frequency in \cite{gwa}. The complexity of wave simulation increases as the third or fourth power of maximum simulation frequency \cite{raghuvanshi2009efficient} and hence it needs to be capped to keep computation time reasonable. GA results are obtained using a geometric simulator that captures both specular and diffuse reflections for the full frequency range (i.e., at a sample rate of 48,000Hz). 
A hybrid scheme (explained in detail in \cite{gwa}) is used to combine RIRs from both  wave and geometric simulations (i.e. hybrid RIRs) at a crossover frequency of 1,400Hz, which yields the \emph{hybrid} RIRs.

\begin{table}[htbp]
\centering
\caption{RIR datasets used for testing our approach.}
\vspace{-3mm}
\label{tab:T60}
\begin{tabular}{|c|c|c|}
\hline
Dataset   & No. of RIRs & $T_{60}$ (in seconds) \\
\hline
MIT       & 270         & 0.41+/-0.48   \\
\hline
VOiCES    & 64          & 3.75+/-0.45   \\
\hline
BUTReverb & 1674        & 1.03+/-0.49   \\
\hline
RWCP Aachen      & 325         & 0.58+/-0.61  \\
\hline
\end{tabular}
\end{table}

\vspace{-2mm}
\subsection{Model Architecture}
\vspace{-2mm}





\begin{table*}[t]
\centering

\caption{Speech dereverberation results on different real world RIR datasets.}
\vspace{-2mm}
\label{tab:results}
\begin{tabular}{|c|c|c|c|c|c|c|c|c|}

\hline
Dataset   &                               No Enhancement                         & WPE  & \multicolumn{3}{c|}{real+synthetic}                  & \multicolumn{3}{c|}{synthetic} \\
\cline{4-9}
          &                                    &      & Hybrid                                & Geo  & FDTD & Hybrid    & Geo     & FDTD      \\
\hline
MIT       &  7.35 & 8.16 & \textbf{8.76} & 8.6  & 7.33 & 7.64      & 7.71    & 5.31    \\
BUTReverb &  3.14 & 3.44 & \textbf{6.7}  & 6.43 & 5.23 & 4.69      & 2.66    & 0.28     \\
VOiCES    &  1.09 & 1.44 & \textbf{4.85} & 3.94 & 2.09 & 2.15      & 1.79    & 1.7      \\
RWCP Aachen      &  5.16 & 5.68 & \textbf{7.21} & 6.85 & 5.34 & 6.08      & 5.23    & 4.69     \\
\hline

\end{tabular}
\end{table*}

We train the SkipConvNet \cite{kothapally2020skipconvnet} model to evaluate the benefits of synthetic RIRs on different datasets. We use the default parameters from their public-domain implementation\footnote{https://github.com/zehuachenImperial/SkipConvNet}. SkipConvNet is a fully-convolutional network that consists of an encoder and a decoder block. The input to the network consists of the log-power spectrum (LPS) of the reverberant speech signal. This model predicts the enhanced LPS, which is combined with the phase of the reverberant signal to generate the enhanced signal. 
The model is optimized using the Mean-Square Error (MSE) loss. More details of the architecture are given in~\cite{kothapally2020skipconvnet}.

\subsection{Experiment Setup}
\vspace{-2mm}
The clean speech signals are obtained from the 100-hour split of the Librispeech \cite{panayotov2015librispeech} dataset. Reverberant signals are generated by convolving the clean speech signals with real and synthetic RIRs. We test our approach on real RIRs from the four different datasets mentioned in Table. \ref{tab:T60}. For each of the test RIR datasets, we sample 50,000 training RIRs for each RIR generation method (hybrid, geometric and FDTD). The RIR sampling procedure is done such that the $T_{60}$ distribution of the sampled RIRs matches that of the corresponding test dataset. 
The model is trained using the ADAM optimizer with an initial learning rate of 0.0001. Early stopping is used to prevent the model from overfitting to synthetically generated RIRs. Two different training runs are carried out. In the first, the training set only consists of synthetically generated RIRs (i.e., only RIRs from the GWA dataset). In the second run, training RIRs consist of real and synthetic RIRs. For the training run with both real and synthetic RIRs, half of the samples in each batch were created from synthetic RIRs; the other half were created from real RIRs. The validation and test sets were generated from a small quantity of real RIRs. 
The real RIRs from the four datasets mentioned in the table are split in the ratio 80:10:10 to generate the train, dev and test splits, respectively.


\section{Results and Analysis}

\subsection{Results}
\vspace{-2mm}
We test our speech dereverberation models on reverberant speech generated using RIRs from four RIR datasets: BUT Reverb dataset~\cite{8717722}, MIT IR dataset~\cite{traer2016statistics}, RWCP Aachen + REVERB RIRs~\cite{ko2017study}, and VOiCES IR dataset~\cite{richey2018voices}. The number of RIRs and the $T_{60}$ distribution of each of these datasets is shown in Table~\ref{tab:T60}.
The main results in terms of accuracy of the speech dereverberation model trained using different sets of RIRs are present in Table \ref{tab:results}. We compute the speech-to-reverberation modulation energy ratio (SRMR) \cite{santos2014improved} as a measure of reverberation. We compute SRMR for two baselines: (a) the reverberant signal without any enhancement applied and (b) the enhanced signal obtained by the Weighted Prediction Error (WPE) \cite{nakatani2010} dereverberation algorithm. 
We obtain the best performance for the hybrid RIR generation approach for the following cases: (a) when training data consists of synthetic RIRs only and (b) when training data consists of synthetic and real RIRs. The addition of real RIRs to the training data significantly improves the SRMR of the enhanced signal. This suggests that there still exists a domain gap between synthetically generated RIRs and real RIRs. However, we  observe that across all RIR datasets and for both real and real+synthetic training runs, the hybrid approach performs better than traditional geometric-based RIR generation methods. The FDTD approach frequently performs worse than the no enhancement baseline, but this is to be expected as the wave solver only simulates frequencies up to 1400 Hz and all higher frequencies in the speech are lost after convolution. In most cases, we observe that both hybrid and geometric approaches perform better than the WPE baseline. For the real+synthetic training run, the hybrid approach shows a relative improvement in the range of $7.5\%$ - $236\%$ across all four test sets when compared to the WPE method. The best improvement in SRMR is observed for the BUTReverb and VOiCES datasets, which consist of RIRs with the largest $T_{60}$ values. Our hybrid approach also results in an average improvement (computed over the four test datasets) of $8.3\%$ for the real+synthetic case and $28.2\%$ for the synthetic only case, as compared to using RIRs generated using only GA methods.
\vspace{-4mm}
\subsection{Analysis of Hybrid vs. Geometric Methods}
\vspace{-2mm}
\begin{table}[]
\caption{We tabulate the relative improvement in SRMR and the fraction of the reverberant signal energy present in the low frequencies for all four datasets.} 
\vspace{-2mm}
\label{tab:tab3}
\begin{tabular}{|c|c|c|}
\hline
\multirow{2}{*}{Dataset} & Rel. Imp. (\%) & Fraction of reverberant         \\
                         & in SRMR        & signal energy \textless 250 Hz \\
\hline
MIT IR                   & 1.86           & 0.17                          \\
\hline
BUTReverb                & 4.19           & 0.4                          \\
\hline
VOiCES                   & 23.09          & 0.66                           \\
\hline
RWCP Aachen     & 4.23           & 0.37    \\
\hline
\end{tabular}
\end{table}

\begin{figure}[t]
\caption{The plot below shows the cumulative signal energy (represented as a fraction of total signal energy) as a function of frequency. The number in brackets in the legend represents the relative improvement offered by hybrid simulation over geometric simulation. } 
\label{fig:energyplot}
\includegraphics[scale=0.4, height=6cm]{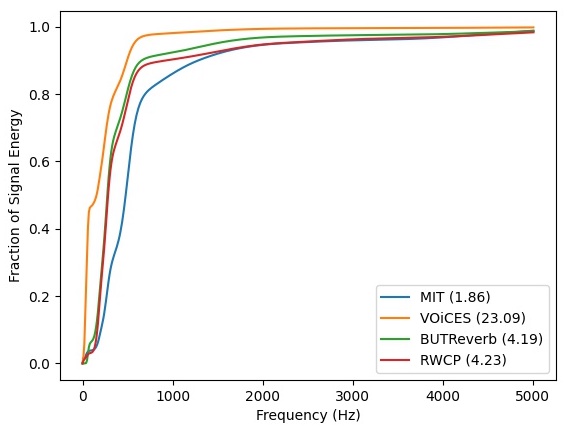}
\centering
\end{figure}

In Table. \ref{tab:tab3}, we tabulate the relative improvement in SRMR and fraction of low-frequency signal energy. For each dataset, we compute the average fraction of signal energy present in the low frequency component ($< 250$ Hz) of the reverberant signal across the entire test set. We choose 250 Hz as the cutoff frequency because the fundamental frequency range of human speech is generally below 250Hz \cite{traunmuller1995frequency}. We observe that this quantity positively correlates with the relative improvement offered by the hybrid method over the geometric method. The Pearson correlation coefficient of these two quantities (present in Table. \ref{tab:tab3}) is 0.9126.
This confirms that more accurate simulation of the low-frequency component of the RIR leads to improved speech dereverberation. Figure. \ref{fig:energyplot} shows the distribution of reverberant signal energy as a function of frequency for all four RIR datasets. From the figure, we can see larger gains from the hybrid approach when the signal energy is more prevalent in lower frequencies.

\section{Conclusion and Future Work}
\vspace{-2mm}
We have demonstrated that accurate wave-based solvers can be used to obtain more accurate simulations of RIRs to train the learning models for speech reverberation. Our proposed approach of augmenting the training process with synthetic RIRs generated from a hybrid combination of geometry-based and wave-based simulations has shown significant improvement in the performance of learning-based speech dereverberation models. We also see a high correlation between the improvement in SRMR with the fraction of the reverberant signal energy at lower frequencies. 



\bibliographystyle{IEEEtran}

\bibliography{mybib}

\end{document}